\documentclass{article}
\usepackage{spconf,amsmath,graphicx}
\usepackage{xcolor}
\usepackage{amssymb}
\usepackage{algorithm,algcompatible,amsmath}
\usepackage{multirow}
\usepackage{siunitx}
\usepackage{booktabs}
\usepackage{algorithm} 
\usepackage{algpseudocode} 
\usepackage{amsthm}
\newtheorem{proposition}{Proposition}
\newtheorem{corollary}{Corollary}
\begin{document}
\ninept

\title{Learning Expanding Graphs for Signal Interpolation}

%
\name{Bishwadeep Das and Elvin Isufi\thanks{Faculty of Electrical Engineering, Mathematics and Computer Science, Delft University of Technology, Delft, The Netherlands. e-mail: \{b.das, e.isufi-1\}@tudelft.nl}
\address{~}}

\maketitle
\begin{abstract}
Performing signal processing over graphs requires knowledge of the underlying fixed topology. However, graphs often grow in size with new nodes appearing over time, whose connectivity is typically unknown; hence, making more challenging the downstream tasks in applications like cold start recommendation. We address such a challenge for signal interpolation at the incoming nodes blind to the topological connectivity of the specific node. Specifically, we propose a stochastic attachment model for incoming nodes parameterized by the attachment probabilities and edge weights. We estimate these parameters in a data-driven fashion by relying only on the attachment behaviour of earlier incoming nodes with the goal of interpolating the signal value. We study the non-convexity of the problem at hand, derive conditions when it can be marginally convexified, and propose an alternating projected descent approach between estimating the attachment probabilities and the edge weights. Numerical experiments with synthetic and real data dealing in cold start collaborative filtering corroborate our findings.
%
\end{abstract}
\begin{keywords}
Incoming nodes, expanded graphs, graph signal interpolation, cold start.
\end{keywords}
\vspace{-3mm}
\section{Introduction}\label{section introduction}
Graph Signal Processing (GSP) leverages the relationships between data points to subsequently process them for a multitude of classical applications \cite{shuman_emerging_2013,ortega_graph_2018}. The graph is commonly considered of fixed size and can have either a fixed \cite{mateos_connecting_2019} or a changing number of edges \cite{kalofolias2017learning}. However, graphs often grow in size with new nodes becoming available continuously \cite{erdos_evolution_1961,barabasi_emergence_1999}. A typical setting is in cold start collaborative filtering. Here, a new item becomes available but we have no information to connect it with the available ones, thereby affecting the subsequent recommendation \cite{liu2020heterogeneous}. Hence, modelling the attachment behaviour of incoming nodes is paramount to interpolate the rating value at this new item \cite{huang_rating_2018,isufi2021accuracy}. {This is relevant not only in recommender systems \cite{liu2020heterogeneous} but also in data privacy over networks \cite{shen_online_2019}, which can manifest in inductive learning on graphs \cite{chen_semi-supervised_2014} where unseen nodes need to be classified.}
\par There exists a set of diverse viewpoints to approach inference of nodal connections \cite{giannakis2018topology}. \textit{Topology Identification} via GSP estimates a static \cite{mateos_connecting_2019,dong2019learning} or a time varying topology \cite{kalofolias2017learning,natali2021online} of fixed size by using different priors such as signal smoothness \cite{kalofolias2016learn}, realizations from diffusion \cite{coutino2020state}, or Gaussian processes \cite{pavez2016generalized}. \textit{Statistical} methods utilize stochastic attachment models based on the existing topology to drive the incoming node attachment. The Erd\H os-R\'enyi (ER) \cite{erdos_evolution_1961} and Barabasi-Albert \cite{barabasi_emergence_1999} models are prime examples of this category, although more complex models also exist \cite{bianconi_competition_2001,zadorozhnyi_growing_2015}. \textit{Link Prediction} approaches infer the existence of unobserved edges between existing nodes, given edges and/or feature information of these nodes \cite{martinez2016survey}. These include probabilistic approaches \cite{clauset2008hierarchical}, similarity-based approaches \cite{liu2008assessing,liu2010link}, and classifier-based approaches \cite{cukierski2011graph}. Some more recent works provide a synthesis between graphs growth models and processing signals over them. The works in \cite{matta2019graph,cirillo2021learning} estimate node connectivity for graphs drawn from ER and Bollob\'as-Riordan models by observing only on a subset of nodes signals that evolve according to an autoregressive process. The works in \cite{shen_online_2019,venkitaraman_recursive_2020} solve regression tasks over expanding graphs but require the connectivity of the incoming nodes.

Altogether, these contributions consider settings where we have a fixed number of nodes, available nodal features, or know the incoming node attachment. In absence of node feature signals and attachment information, both the GSP and the link prediction approaches become inapplicable. We may in these cases rely on stochastic models but they detach the connectivity modelling from the processing task, which results in sub-optimal performance. Likewise, we can also adapt the approaches in \cite{matta2019graph,cirillo2021learning} to account for incoming nodes but the considered growth and signal models do not always hold. 
\par To overcome these limitations, we propose a stochastic attachment model for arbitrary graphs and utilize the information only on the existing graph to interpolate the signal value at the incoming node.
The proposed model is parameterized by the attachment probabilities and the edge weights of the incoming node. These parameters are estimated in a data- and task-driven fashion from a training set of earlier incoming nodes. To estimate the parameters, we solve an empirical risk minimization problem by minimizing the signal reconstruction mean squared error on the incoming node regularized to the connectivity pattern. We study the convexity of the problem and provide an alternating projected descent algorithm to solve it. Finally, we corroborate the proposed model and compare it with baselines on synthetic and real applications dealing with the cold start problem in collaborative filtering.
\vspace{-0.35cm}
\section{Problem Formulation}\label{Section Problem Formulation}
Consider a graph $\mathcal{G}=(\mathcal{V},\mathcal{E})$ of $N$ nodes in set $\mathcal{V}=\{v_1,\ldots,v_N\}$ and $E$ edges in set $\mathcal{E}\subseteq\mathcal{V}\times\mathcal{V}$. Let $\mathbf{A}$ be the graph adjacency matrix such that $A_{ij}\neq0$ only if $(v_i, v_j) \in \mathcal{E}$. An incoming node $v_+$ connects to $\mathcal{G}$ and forms a directed expanded graph $\mathcal{G}_+=(\mathcal{V}_+,\mathcal{E}_+)$ with node set $\mathcal{V}_+=\mathcal{V}\cup v_+$ and edge set $\mathcal{E}_+=\mathcal{E}\cup(v_+,v_i)$ for all new directed edges $(v_+,v_i)$ landing at $v_+$. 
We represent the attachment pattern of node $v_+$ by $\mathbf{a}_+\in\mathbb{R}^{N}$ where $[\mathbf{a}_+]_i=w_i$ is the weight of edge $(v_+,v_i)$. The $(N+1) \times (N+1)$ adjacency matrix of graph $\mathcal{G}_+$ is
\begin{equation}\label{A_+ equation}
\mathbf{A}_+=\begin{bmatrix}
\mathbf{A} & \mathbf{0}_{+} \\
\mathbf{a}_{+}^{\top} & 0 \\
\end{bmatrix}
\end{equation}
in which the last row and column represent the connectivity of $v_+$.\footnote{We consider for simplicity of exposition the attachment of a single node. For multiple nodes, $\mathbf{a}_+$ becomes a matrix having a column per new node.}
Node $v_+$ connects to any existing node $v_i \in \mathcal{V}$ independently with probability $p_i$. We collect the probabilities of attachment in vector $\mathbf{p}=[p_1,\ldots,p_N]^{\top}$. Then, the attachment pattern $\mathbf{a}_+$ is a random vector with each entry being an independent Bernoulli random variable weighted by scalar $w_i$. I.e., the $i$th element of $\mathbf{a}_+$ is
\begin{equation}
     [\mathbf{a}_+]_i= \begin{cases}
               w_i              & \text{with probability}  \hspace{1mm} p_i\\
               0              & \text{with probability} \hspace{1mm} (1-p_i)\\
           \end{cases}  
\end{equation}for $i=1,\ldots,N$. The expectation of $\mathbf{a}_+$ is $\mathbb{E}[\mathbf{a}_+]=\mathbf{p}\circ\mathbf{w}$ where vector $\mathbf{w}=[w_1,\ldots,w_N]^{\top}$ contains the weights of all edges $(v_+,v_i)$. Likewise, the variance of $[\mathbf{a}_+]_i$ is $\text{var}([\mathbf{a}_+]_i)=w_i^2p_i(1-p_i)$ and the respective covariance matrix is
\begin{equation}\label{covariance matrix formula}
    \boldsymbol{\Sigma}_+=\text{diag}(\mathbf{w}^{\circ2}\circ\mathbf{p}\circ(\mathbf{1}-\mathbf{p}))
\end{equation}
where $\mathbf{a}^{\circ k} = \mathbf{a}\circ \ldots \circ \mathbf{a}$ is the Hadamard product of $\mathbf{a}$ with itself $k$ times. The expected adjacency matrix of $\mathcal{G}_+$ is
\begin{equation}\label{expected A}
    \mathbb{E}[\mathbf{A}_+]=\begin{bmatrix}
\mathbf{A} & \mathbf{0}_+ \\
(\mathbf{p}\circ\mathbf{w})^{\top} & 0 \\
\end{bmatrix}.
\end{equation}

Let $\mathbf{x}=[x_1,\ldots,x_N]^{\top}$ be the graph signal on $\mathcal{G}$ with $x_i$ the signal at node $v_i$. Processing this signal by accounting for its coupling with $\mathcal{G}$ is key to several network data tasks \cite{shuman_emerging_2013}. E.g., in item-item collaborative filtering, the signal collects the ratings a specific user has provided to the existing items \cite{huang_rating_2018}. When a new item node $v_+$ becomes available, the task is to predict the signal rating value $x_+$ at this node. To solve such a task without knowing the exact connectivity of $v_+$, we rely on the stochastic models governed by $\mathbf{p}$ and $\mathbf{w}$, which in turn are unknown.

To identify a task-specific connectivity for the incoming nodes, we merge data-driven solutions with the above generic statistical model. Given graph $\mathcal{G}$ and a training set of attachment patterns for incoming nodes $\mathcal{T}=\{(v_{t+},x_{t+},\mathbf{a}_{t+},\mathbf{b}_{t+})\}_t$ we infer the attachment probability vector $\mathbf{p}$ and weight vector $\mathbf{w}$ in an empirical risk minimization fashion. Each element in $\mathcal{T}$ comprises an incoming node 
$v_{t+}$, the signal at this node $x_{t+}$, the attachment vector $\mathbf{a}_{t+}$, and its binary attachment pattern $\mathbf{b}_{t+}$ --for a recommender system with cold start, we build an item-item graph $\mathcal{G}$ and treat some items as cold-starters whose ratings and connectivity are known -- and define a task-specific loss $f_{\mathcal{T}}(\mathbf{p},\mathbf{w},\mathbf{a}_{t+},\mathbf{x}_{t+})$ measuring the signal interpolation (rating) performance. Specifically, we solve the statistical optimization problem
\begin{align}\label{pbasic}
\begin{split}
    &\underset{\mathbf{p},\mathbf{w}}{\text{min.}} \mathbb{E}\big[f_{\mathcal{T}}(\mathbf{p},\mathbf{w},\mathbf{a}_{t+},\mathbf{x}_{t+})\big]+g_{\mathcal{T}}(\mathbf{p},\mathbf{b}_{t+})
    +h_{\mathcal{T}}(\mathbf{w},\mathbf{a}_{t+})\\
    &\text{subject to} \quad  \mathbf{p}\in[0,1]^N, \mathbf{w}\in\mathcal{W}\hspace{2mm}
\end{split}
\end{align}
where $g_{\mathcal{T}}(\mathbf{p},\mathbf{b}_{t+})$ and $h_{\mathcal{T}}(\mathbf{w},\mathbf{a}_{t+})$ are regularizers and set $\mathcal{W}$ constraints the edge-weights, e.g., non-negative or finite.

\section{Signal Interpolation on Incoming nodes }\label{Section Method}
We measure the signal interpolation performance via the mean square error (MSE) and use graph filters \cite{sandryhaila_discrete_2013,coutino_advances_2019} to diffuse the signal over $\mathcal{G_+}$. Graph filters are well-established local operators for processing graph data. They combine successive shift operations over the topology and have found applications in a variety of domains \cite{ortega2018graph}
Consider the expanded graph signal $\mathbf{x}_+=[\mathbf{x}^{\top},0]^{\top}$, where zero is the signal value at node $v_+$. The output $\mathbf{y_+}$ of an order $L$ graph filter is 
\vspace{-2mm}
\begin{equation}\label{FIR GC}
    \mathbf{y}_+=\sum_{l=1}^{L}h_l\mathbf{A}_+^l\mathbf{x}_+
\end{equation}
where $\mathbf{h}=[h_1,\ldots,h_L]^{\top}$ are the filter coefficients. Nodes up to the $L$-hop neighborhood of $v_+$ contribute to its interpolated signal. Note that in \eqref{FIR GC}, we ignore $l = 0$ because it does not contribute to the output $[\mathbf{y}_+]_{N+1}$ at $v_+$. Given the percolated signal $[\mathbf{y}_+]_{N+1}$, the following proposition quantifies the signal interpolation MSE as a function of model parameters $\mathbf{p}$ and $\mathbf{w}$. 
\begin{proposition}\label{prop_MSE}
\textit{Given $\mathcal{G}=\{\mathcal{V},\mathcal{E}\}$ with adjacency matrix $\mathbf{A}$ and signal $\mathbf{x}$ and let $\mathbf{A}_x=[\mathbf{x},\ldots,\mathbf{A}^{L-1}\mathbf{x}]$. Given also an incoming node $v_+$ with true signal $x^{\star}_+$ attaching to $\mathcal{G}$ with probabilities $\mathbf{p}$ and edge weights $\mathbf{w}$, forming graph $\mathcal{G}_+$ with the expanded adjacency matrix $\mathbf{A}_+$ $[cf. \eqref{A_+ equation}]$. The MSE of the interpolated signal $\mathbf{y}_+$ at node $v_+$ by an order $L$ graph filter $[cf.\eqref{FIR GC}]$ is}
\begin{equation}\label{MSE formula}
    \textnormal{MSE}(\mathbf{p},\mathbf{w})= ||(\mathbf{w}\circ\mathbf{p})^{\top}\mathbf{A}_x\mathbf{h}-x^{\star}_+||^2_2+\mathbf{h}^{\top} \mathbf{A}_x^{\top}\boldsymbol{\Sigma}_+\mathbf{A}_x\mathbf{h}
\end{equation}
\end{proposition}

\noindent\textit{Proof.} See appendix.\qed

Besides quantifying the MSE, Proposition \ref{prop_MSE} provides also insights on the role of parameters $\mathbf{p}$ and $\mathbf{w}$. The first term on the RHS of \eqref{MSE formula} captures the model bias w.r.t. the true signal $x^\star_+$. The prediction output is the dot product between the filtered output of $\mathbf{x}$ over $\mathcal{G}$, $\mathbf{A}_x\mathbf{h}$ with the expected attachment vector $\mathbf{w}\circ\mathbf{p}$. Minimizing the bias implies selecting a pair $(\mathbf{p}$, $\mathbf{w})$ that combines the signal at each $v\in\mathcal{V}$ to match $x^{\star}_+$.
The second term $\mathbf{h}^{\top} \mathbf{A}_x^{\top}{\boldsymbol{\Sigma}_+}\mathbf{A}_x\mathbf{h}=\|\mathbf{A}_x\mathbf{h} \|_{\boldsymbol{\Sigma}_+}^2$ is the squared norm of the filtered signal weighted by the attachment variances. Minimizing this term might give trivial solutions such as $\mathbf{p}=\mathbf{1}$ and $\mathbf{p}=\mathbf{0}$. Thus, regularizers are needed for $\mathbf{p}$ and $\mathbf{w}$. We also remark that in \eqref{MSE formula} the $L$th shift $\mathbf{A}^L\mathbf{x}_+$ does not contribute to the MSE because of the structure of matrix $\mathbf{A}_+$ in \eqref{A_+ equation}.

\par\smallskip\noindent\textbf{Optimization Problem.} With this in place, we can formulate problem \eqref{pbasic} as 
\vspace{-0.2cm}
\begin{align}\label{Interpolation main problem}
\begin{split}
    &\underset{\mathbf{p},\mathbf{w}}{\text{min.}} \quad \text{MSE}_{\mathcal{T}}(\mathbf{p},\mathbf{w})
    +\sum_{t=1}^{|\mathcal{T}|}\bigg(\mu_p||\mathbf{p}-\mathbf{b}_{t+}||_q^q+\mu_w||\mathbf{w}-\mathbf{a}_{t+}||_q^q\bigg)\\
    &\text{subject to} \quad  \mathbf{p}\in[0,1]^N, \mathbf{w}\in\mathcal{W}
    \end{split}
\end{align}
where $\text{MSE}_{\mathcal{T}}(\mathbf{p},\mathbf{w})$ is the empirical MSE over the training set $\mathcal{T}$, $\sum_{t=1}^{|\mathcal{T}|}||\mathbf{p}-\mathbf{b}_{t+}||_q^q$ and $\sum_{t=1}^{|\mathcal{T}|}||\mathbf{w}-\mathbf{a}_{t+}||_q^q$ are regularizers with weights $\mu_w>0$, $\mu_p>0$. respectively and $q\in\{1,2\}$. 
\begin{algorithm}[!t]
	\caption{Alternating projected gradient descent for \eqref{Interpolation main problem}.}
	\begin{algorithmic}[1]
	\State \textbf{Input:} Graph $\mathcal{G}$, training set $\mathcal{T}$, graph signal $\mathbf{x}$, adjacency matrix $\mathbf{A}$, number of iterations $K$, cost $C$, learning rates $\lambda_p,\lambda_w$.
	\State \textbf{Initialization}: $\mathbf{p}=\mathbf{p}^0$, $\mathbf{w}=\mathbf{w}^0$ randomly, $k=0$. 
	\For {$k \leq K$}
	\State\text{$\mathbf{p}$ gradient}: $\tilde{\mathbf{p}}^{k+1}=\mathbf{p}^{k}-\lambda_p\nabla_{\mathbf{p}}C(\mathbf{p}^k,\mathbf{w}^k)$;
	\State\text{Projection}\label{step5}: $\mathbf{p}^{k+1}=\underset{[0,1]^N}{\Pi}(\tilde{\mathbf{p}}^{k+1})$;
	\State\text{$\mathbf{w}$ gradient}: $\mathbf{w}^{k+1}=\mathbf{w}^{k}-\lambda_w\nabla_{w}C(\mathbf{p}^{k+1},\mathbf{w}^{k})$;
	\State\text{Projection}: $\mathbf{w}^{k+1}=\underset{\mathcal{W}}{\Pi}(\tilde{\mathbf{w}}^{k+1})$;
	\EndFor
	\end{algorithmic} 
	\label{Algorithm MSE}
\end{algorithm}
Problem \eqref{Interpolation main problem} is non-convex in $\mathbf{w}$ and $\mathbf{p}$; it is marginally convex in $\mathbf{w}$ but not always in $\mathbf{p}$ due to the variance term in \eqref{MSE formula}. We solve \eqref{Interpolation main problem} with alternating projected gradient descent. Algorithm \ref{Algorithm MSE} summarizes the main steps. The gradients of the cost $C(\mathbf{p},\mathbf{w})$ in \eqref{Interpolation main problem} w.r.t. $\mathbf{p}$ and  $\mathbf{w}$ for $q=2$ are 
\vspace{-3.8mm}
\begin{align}\label{der p MSE}
\begin{split}
    & \nabla_{p}C(\mathbf{p},\mathbf{w})=2\sum_{t=1}^{|\mathcal{T}|}((\mathbf{w}\circ\mathbf{p})^{\top}\mathbf{A}_x\mathbf{h}-x_{t+})(\mathbf{w}\circ\mathbf{A}_x\mathbf{h})\\
    &+|\mathcal{T}|(\mathbf{A}_x\mathbf{h})^{\circ2}\circ(\mathbf{w}^{\circ2})\circ(\mathbf{1}-2\mathbf{p})+2\mu_p\sum_{t=1}^{|\mathcal{T}|}(\mathbf{p}-\mathbf{b}_{t+})
\end{split}
\end{align}
\begin{align}\label{der w MSE}
\begin{split}
    & \nabla_{w}C(\mathbf{p},\mathbf{w})=2\sum_{t=1}^{|\mathcal{T}|}((\mathbf{w}\circ\mathbf{p})^{\top}\mathbf{A}_x\mathbf{h}-x_{t+})(\mathbf{p}\circ\mathbf{A}_x\mathbf{h})\\
    &+2|\mathcal{T}|(\mathbf{A}_x\mathbf{h})^{\circ2}\circ\mathbf{w}\circ\mathbf{p}\circ(\mathbf{1}-\mathbf{p})+2\mu_w\sum_{t=1}^{|\mathcal{T}|}(\mathbf{w}-\mathbf{a}_{t+})
\end{split}.    
\end{align}
Instead, for $q = 1$, we replace terms $2\mu_p(\mathbf{p}-\mathbf{b}_{t+})$ and $2\mu_w(\mathbf{w}-\mathbf{a}_{t+})$ with $\text{sign}(\mathbf{p}-\mathbf{b}_{t+})$ and $\text{sign}(\mathbf{w}-\mathbf{a}_{t+})$  respectively. We have observed that norm one regularizers are more applicable to $\mathbf{p}$ than to $\mathbf{w}$ because some weights would be zero even if a node attaches with a high probability. While we can use Algorithm \ref{Algorithm MSE} to solve the non-convex case of problem \eqref{Interpolation main problem}, the following corollary provides a sufficient condition for problem \eqref{Interpolation main problem} to be marginally convex also in $\mathbf{p}$. The proposed method has a complexity of order $\mathcal{O}(LE)$, where $L$ is the filter order and $E$ the number of edges in the existing graph.
\begin{corollary}
\label{marginal convexity}
\textit{The cost in Problem \eqref{Interpolation main problem} is marginally convex in $\mathbf{p}$ if the regularization weight $\mu_p > 0$ satisfies}
\begin{align}\label{mu_p condition}
     \mu_p\geq w_h^2\underset{i\in\{1,\ldots,N\}}{\textnormal{max}}([\mathbf{A}_x\mathbf{h}]_i)^2-||\mathbf{w}\circ\mathbf{A}_x\mathbf{h}||_2^2.
\end{align}
\end{corollary}
\noindent\textit{Proof.} See appendix. \qed

While guaranteeing convexity, condition \eqref{mu_p condition} may lead to an optimum that is worse than the local optima of its non-convex counterpart due to a greater focus on the training attachment patterns than on the task-specific cost. We shall corroborate this next.
\vspace{-2mm}
\section{Numerical Results}\label{Section Evluation}
In this section, we evaluate our approach on synthetic and real data. For comparison, we consider:
    $i)$ \textit{uniform attachment}: the incoming node attaches uniformly, i.e., $
    \mathbf{p}_{\text{rd}}=\frac{1}{N}\mathbf{1}$;
    $ii)$ \textit{preferential attachment}: the incoming node attaches with probability vector
    $\mathbf{p}_{\text{pf}}={\mathbf{d}}/{\mathbf{1}^{\top}\mathbf{d}}$ where $\mathbf{d}$ is the degree vector;
    $iii)$ \textit{training attachment only}: a data-driven rule where we rely only on the attachment patterns available during training to build $\mathbf{p}_{g}=\frac{1}{|\mathcal{T}|}\sum_{t=1}^{|\mathcal{T}|}\mathbf{b}_{t+}$ and $\mathbf{w}_{g}=\frac{1}{|\mathcal{T}|}\sum_{t=1}^{|\mathcal{T}|}\mathbf{a}_{t+}$, i.e., we ignore the MSE costs. The first two serve as baselines to assess how the propsoed data-driven stochastic model compares with conventional statistical models, while the latter is considered to assess the importance of the task-specific cost.
\vspace{-4mm}
\begin{figure}[t]
\centering
\includegraphics[trim=0 290 30 280,clip,width=0.24\textwidth]{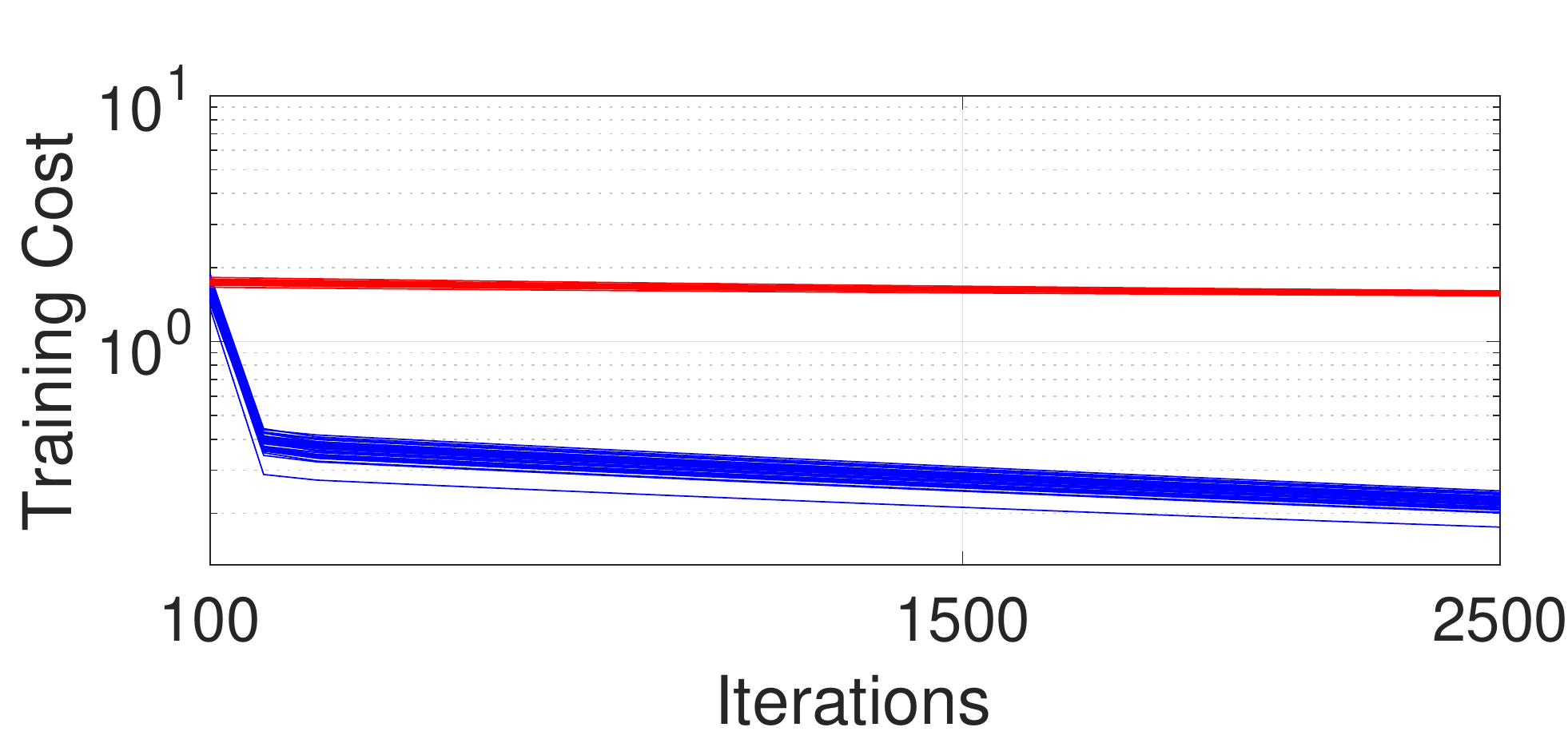}%
{\includegraphics[trim=32 300 30 300,clip,width=0.24\textwidth]{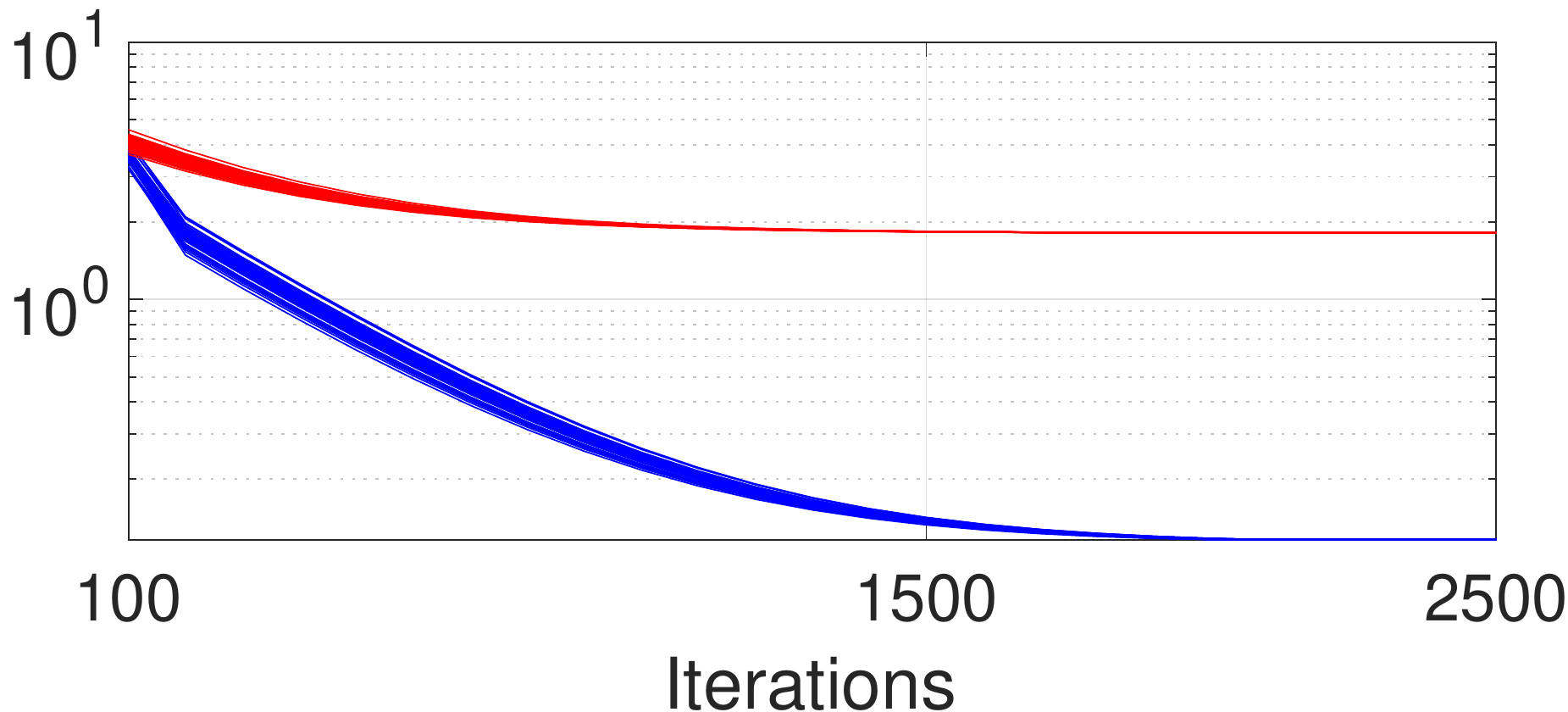}}%
\caption{Convergence of the training cost for Algorithm~\ref{Algorithm MSE} solving Problem~\eqref{Interpolation main problem} being non-convex (blue) and marginally convex (red) [cf. \eqref{mu_p condition}]. (Left) ER graph model; (Right) BA graph model.
%
}
\label{training}
\end{figure}
\begin{table}[t]
\centering
\label{MSE comparison}
\caption{Averaged MSE and its standard deviation (Std) for synthetic data. (Upper) comparison with the uniformly random and preferential attachment; (Lower) joint training vs. individual training.}
\begin{tabular}{|l|l l l|l l l|} 
\hline
 & \multicolumn{3}{|c|}{Erd\H os-R\'enyi} & \multicolumn{3}{|c|}{Barabasi-Albert}
\\\hline
\footnotesize{\textbf{}} & Prop. & Pref. & Rand. & Prop. & Pref. & Rand. 
\\\hline 
 \footnotesize{\textbf{MSE}} & \bf{0.03}  & 0.06 & 0.06  & \bf{0.05} & 0.1 & 0.08 \\\hline
 \footnotesize{\textbf{Std.}} & \bf{0.003} & 0.003 & 0.003 & \bf{0.006} & 0.006 & 0.006 \\\hline\hline
\footnotesize{\textbf{}} & $\mathbf{p}$,$\mathbf{w}$ & only $\mathbf{p}$& only $\mathbf{w}$ & $\mathbf{p}$,$\mathbf{w}$ & only $\mathbf{p}$ & only $\mathbf{w}$ \\\hline
 \footnotesize{\textbf{MSE}} & \bf{0.03} & 0.07 & 0.039 & \bf{0.05} & 0.11  & 0.05 \\ \hline
 \footnotesize{\textbf{Std.}} & \bf{0.003}  & 0.003 & 0.003 & \bf{0.006} & 0.005 & 0.006 \\ \hline
\end{tabular}
\label{mse table}
\end{table}

\subsection{Synthetic Data}

For the synthetic experiments, we started with two undirected and unweighted random graphs following the Erd\H os-R\'enyi and Barabasi-Albert models each having $N=100$ nodes. The respective graph signals were formed by randomly combining the first $30$ eigenvectors of the corresponding graph Laplacian matrices. We then normalized the signal to have zero mean. The edge formation probabilities for these graphs and for the respective incoming nodes were set as $\mathbf{p}_{\text{rd}}$ and $\mathbf{p}_{\text{pf}}$ for the ER and BA graphs, respectively. We used a filter of order $L=3$ with coefficients $h_l=\alpha^l$ and $\alpha=0.3$ to diffuse the signal. The training set comprises $1000$ data-points with a 800-200 train-test split and we selected $\mu_p$, $\mu_w$  via ten-fold cross-validation from $[10^{-5},10^{0}]$. The learning rates $\lambda_p$, $\lambda_w$ were fixed to $10^{-5}$. We average the reconstruction MSE over $100$ realizations per test node error and $100$ train-test splits for a total of $10^4$ runs.
\par First, we assess the convergence of Algorithm~\ref{Algorithm MSE} under both the convex and non-convex settings. For the marginally convex condition also in $\mathbf{p}$ we set $\mu_p=30$ to satisfy the convexity criterion in \eqref{mu_p condition}. Fig. \ref{training} shows the training costs as a function of the number of iterations for $50$ random initializations. We observe that the proposed approach converges always to local minima and it reaches a lower value for the non-convex case (blue) compared with the convex one (red) for all initialization. This is because a higher weight $\mu_p$ on the regularizer results in $\mathbf{p}$ adapting more to fit the training attachments $\mathbf{b}_{t+}$ rather than interpolating the signal, ultimately underfitting.

Next, we compare the interpolation MSE with the baselines for the ER ($\mu_p=1, \mu_w=1$) and BA ($\mu_p=1,\mu_w=0.1$). In the upper part of Table \ref{mse table}, we see the proposed approach outperforms alternatives in both settings. To further investigate the role of $\mathbf{p}$ and $\mathbf{w}$, we also train the proposed method for each of $\mathbf{p}$ and $\mathbf{w}$ while keeping the other fixed. The lower half of Table \ref{mse table} suggests training only $\mathbf{w}$ provides a performance comparable to the joint training, which shows that the proposed approach still reaches the optimal value even without knowing the true attachment. However, training only $\mathbf{p}$ degrades the performance appreciably. This is because when $\mathbf{p}$ is known, we train on $\mathbf{w}$ with a convex cost and reach the global minima, as opposed to training only for $\mathbf{p}$ over a non-convex cost.

\subsection{Cold start Collaborative Filtering}

\begin{figure*}[t]
\centering
\includegraphics[trim=30 230 50 230,clip,width=0.32\textwidth]{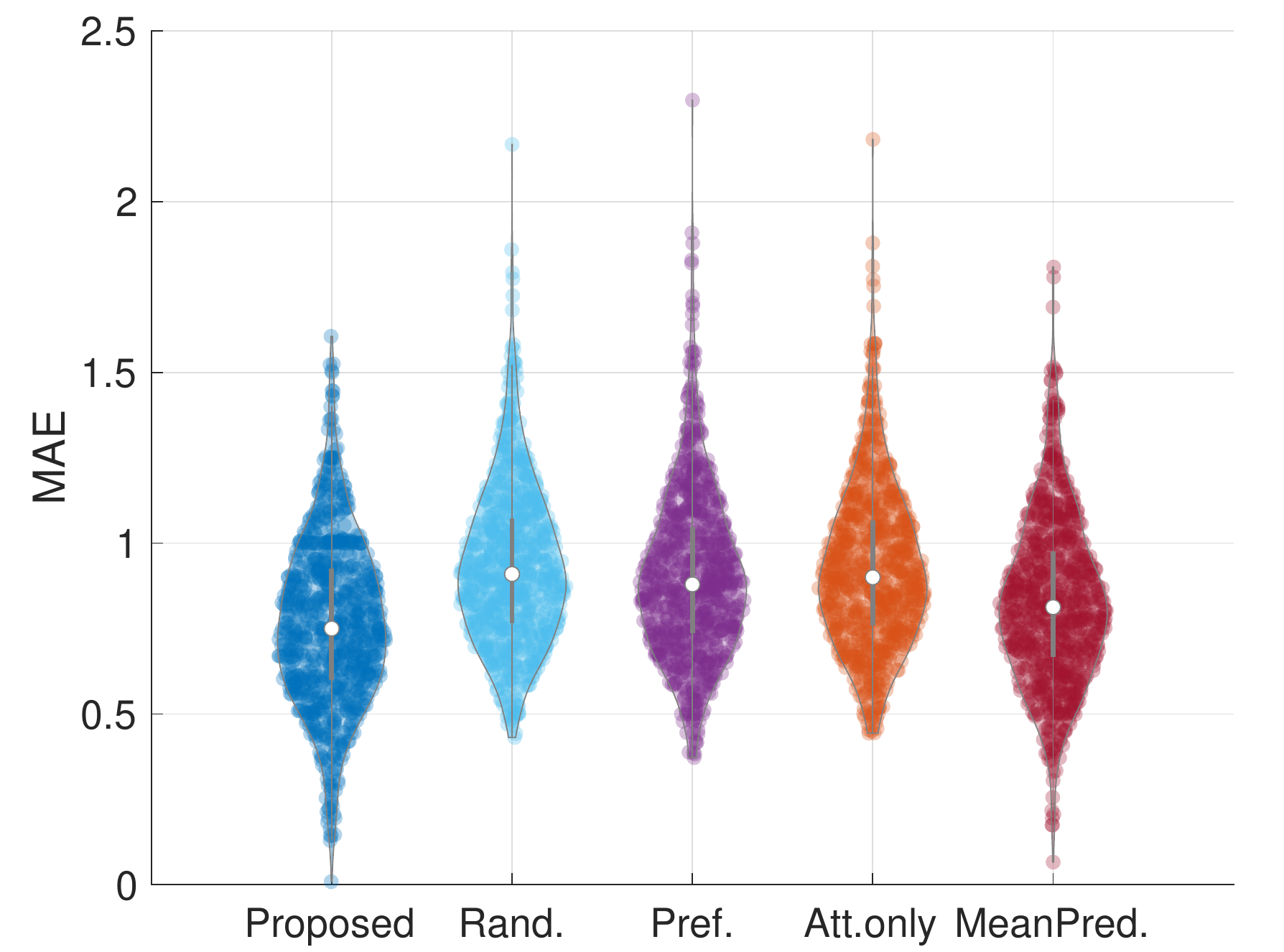}%
{\includegraphics[trim=30 230 50 230,clip,width=0.32\textwidth]{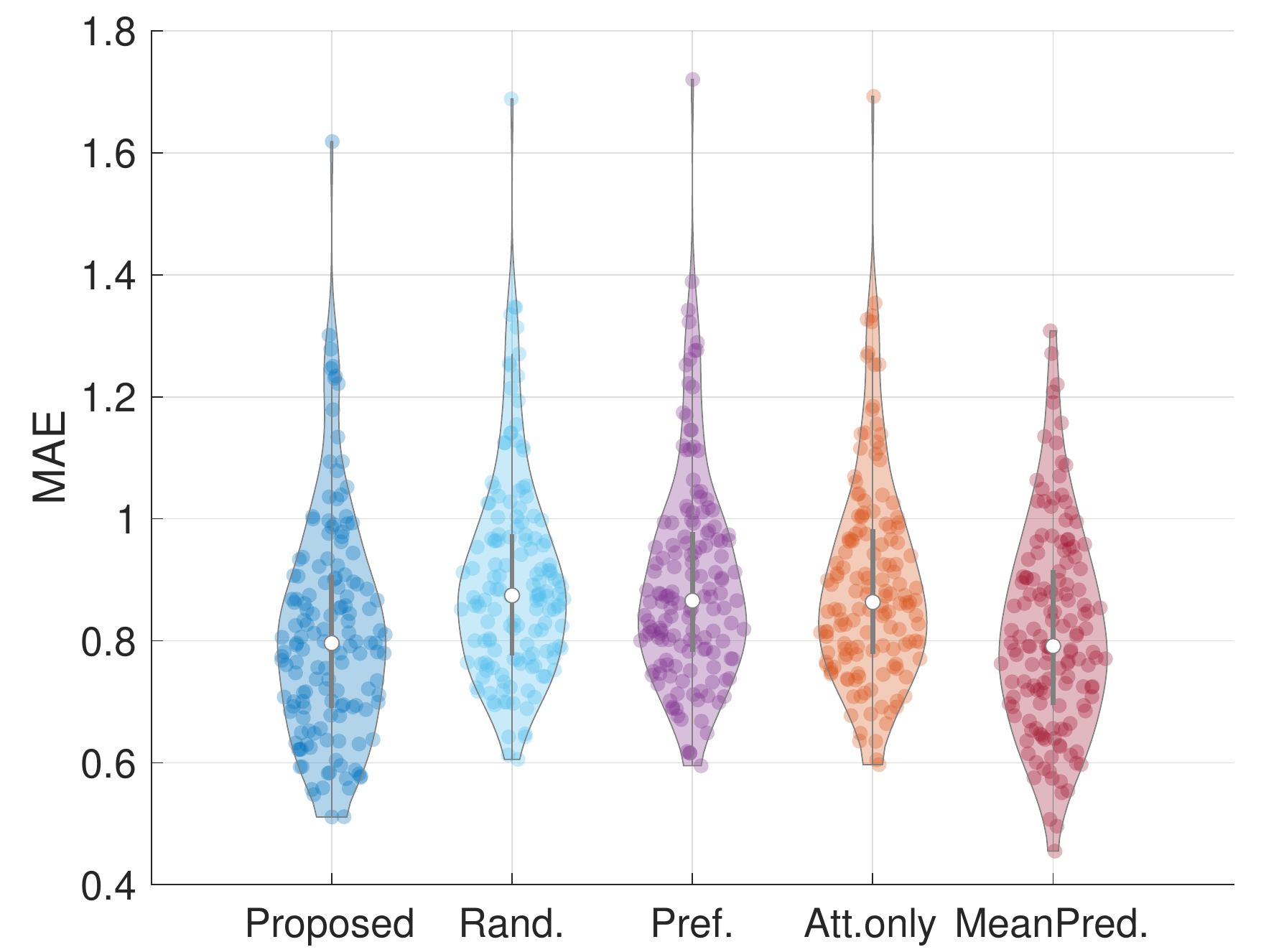}}%
{\includegraphics[trim=30 230 50 230,clip,width=0.32\textwidth]{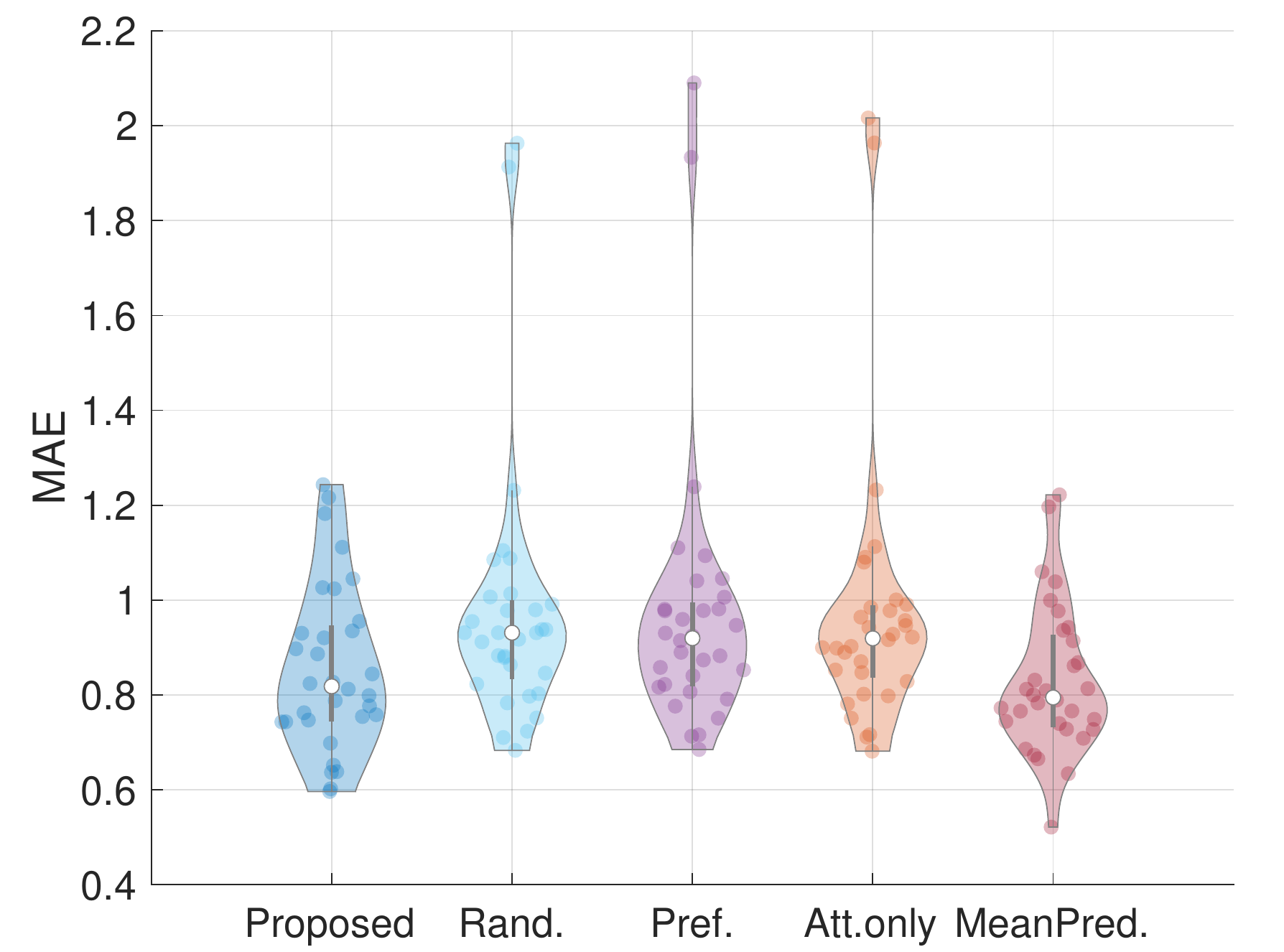}}%
\caption{Mean absolute error (MAE) violin plots for different methods and different rating densities. (Left) low ratings - proposed does best $(0.75\pm0.24)$, followed by mean $(0.81\pm0.24)$; (Centre) medium ratings - proposed and mean $(0.79\pm0.16)$ are tied; (Right) high ratings - mean does best $(0.79\pm0.15)$, followed by proposed $(0.81\pm0.17)$.}
\label{violin}
\end{figure*}

We now use the proposed method for rating prediction in cold start item-item collaborative filtering. We considered the Movielens 100K data-set \cite{harper2015movielens} and removed all entities having less than 10 ratings leading to 943 users and 1152 items. For the cold start experiment, we start with a set of items with known ratings and use those to predict the ratings for unseen items for each user.

We built an item-item $35$ nearest neighbour directed graph comprising $50$ node items following \cite{huang_rating_2018}. The remaining items are divided into 700 for training and 402 for testing. To predict the ratings we used an order five graph filter, which coefficients are estimated to predict the ratings on the initial user-item set as in \cite{huang_rating_2018}. For our problem, we imposed a sparsity constraint on $\mathbf{p}$ and an $\ell_2-$norm on $\mathbf{w}$. Algorithm \ref{Algorithm MSE} is run for $2000$ iterations with learning rates $\lambda_p,\lambda_w=10^{-4}$. We predict ratings on the test items for each user individually. We considered the Mean Absolute Error (MAE) as the evaluation criterion averaged over $100$ samples drawn from $\mathbf{p}$. 


In Fig. \ref{violin} we show the obtained results, where in addition to the former baselines we also considered the user-based mean prediction. To better interpret the results, we divided users into three categories: $i)$ \textit{low} containing users with less than $100$ interactions; $ii)$ \textit{medium} containing users with $100$ to $200$ interactions, and $iii)$ \textit{high} containing users with more than $200$ interactions. We observe that for users with low interactions the proposed approach achieves the lowest median error and personalizes the recommendations. For the medium and high interactions available, the proposed method performs better than the other attachment metrics. This shows the benefits of learning a graph and data-specific attachment pattern suited to a task. Second, for high rating users, the three starting baselines have a long tail in the MAE, whereas ours is more robust. The mean prediction method performs equally well in the high setting because in this case we have more ratings available and the cold start problem is less relevant. Instead, the proposed approach yield benefits in more data scarcity settings which pose the real challenge in recommender systems.

\vspace{-0.5cm}
\section{Conclusion}\label{Section Conclusion}

We proposed a data-driven attachment model for signal interpolation at the incoming nodes. The proposed model is characterized by probabilities of attachment and weights which we used training data to estimate for signal interpolation. We formulated a stochastic optimization problem w.r.t. the attachment parameters and used an alternating projected descent to solve it. We provided conditions when we can relax our interpolation requirements to make the convex marginally convex in both variables. The proposed approach outperforms related statistical models for interpolation over both synthetic and real experiments on cold start collaborative filtering. Future work will consider extending this approach to handle a sequence of nodes.


\section{Appendix}\label{Appendix MSE}
\noindent\textit{Proof of Proposition 1:}
The interpolation MSE is $\mathbb{E}[([\mathbf{y}_+]_{N+1}-x^{\star}_+)^2]$ with  the output at the incoming node. To express $[\mathbf{y}_+]_{N+1}$ in the model parameters, we consider the $l$th power of the adjacency matrix
\begin{equation}
\mathbf{A}_+^l=\begin{bmatrix}
\mathbf{A}^l & \mathbf{0} \\
\mathbf{a}_{+}^{\top}\mathbf{A}^{l-1} & 0 \\
\end{bmatrix}
\end{equation} 
and substitute it in the output $\mathbf{y}_+=\sum_{l=1}^{L}\mathbf{A}_+^l\mathbf{x}_+$ [cf. \eqref{FIR GC}]. The output at the incoming node is thus $[\mathbf{y}_+]_{N+1}=\mathbf{a}_+^{\top}\sum_{l=1}^{L}\mathbf{A}_+^{l-1}\mathbf{x}=\mathbf{a}_+^{\top}\mathbf{A}_x\mathbf{h}$ where $\mathbf{A}_x=[\mathbf{x},\ldots,\mathbf{A}^{L-1}\mathbf{x}]$ and $\mathbf{h}=[h_1,\ldots,h_L]^{\top}$. The MSE is thus
\begin{equation}
\text{MSE}(\mathbf{p},\mathbf{w})=\mathbb{E}[(\mathbf{a}_+^{\top}\mathbf{A}_x\mathbf{h}-x^{\star}_+)^2].
\end{equation}
We add and subtract $(\mathbf{w}\circ\mathbf{p})^{\top}\mathbf{A}_x\mathbf{h}$ within the expectation and get
\begin{align}
    &\text{MSE}(\mathbf{p},\mathbf{w})=\mathbb{E}[((\mathbf{a}_+-\mathbf{w}\circ\mathbf{p})^{\top}\mathbf{A}_x\mathbf{h}+(\mathbf{w}\circ\mathbf{p})^{\top}\mathbf{A}_x\mathbf{h}-x^{\star}_+)^2]
\end{align}where the RHS upon expanding becomes
\begin{align}
\begin{split}
    & \mathbb{E}[(\mathbf{a}_+^{\top}\mathbf{A}_x\mathbf{h}-(\mathbf{w}\circ\mathbf{p})^{\top}\mathbf{A}_x\mathbf{h})^2]+\mathbb{E}[(\mathbf{w}\circ\mathbf{p})^{\top}\mathbf{A}_x\mathbf{h}-x^{\star}_+)^2]\\
    &+2\mathbb{E}[(\mathbf{a}_+^{\top}\mathbf{A}_x\mathbf{h}-(\mathbf{w}\circ\mathbf{p})^{\top}\mathbf{A}_x\mathbf{h})((\mathbf{w}\circ\mathbf{p})^{\top}\mathbf{A}_x\mathbf{h}-x^{\star}_+)].
\end{split}
\end{align}
In the first term, we expand the square, factor $\mathbf{A}_x\mathbf{h}$, and take the expectation inside to get $(\mathbf{A}_x\mathbf{h})^{\top}\mathbb{E}[(\mathbf{a}_+-\mathbf{w}\circ\mathbf{p})(\mathbf{a}_+-\mathbf{w}\circ\mathbf{p})^{\top}]\mathbf{A}_x\mathbf{h}$ which writes as $(\mathbf{A}_x\mathbf{h})^{\top}\boldsymbol{\Sigma}_+\mathbf{A}_x\mathbf{h}$ [cf. \eqref{covariance matrix formula}]. The second term is deterministic, thus we can drop the expectation. The third term instead is zero because $\mathbb{E}[\mathbf{a}_+]=\mathbf{w}\circ\mathbf{p}$. Combining these, we get \eqref{MSE formula}. \hfill\qed

\smallskip
\noindent\textit{Proof of Corollary 1}:\label{marginal}
\noindent 
To get marginal convexity, we check when the Hessian of the cost in \eqref{Interpolation main problem} is positive semi-definite. The derivative of \eqref{Interpolation main problem} w.r.t. $\mathbf{p}$ is shown in \eqref{der p MSE}. The Hessian in $\mathbf{p}$ is
\begin{align}\label{dzn}
\begin{split}
    \nabla_p^2 C(\mathbf{p},\mathbf{w})&\!=\!2(\!\mathbf{w}\!\circ\!\mathbf{A}_x\mathbf{h}\!)\!(\!\mathbf{w}\!\circ\!\mathbf{A}_x\mathbf{h}\!)^{\top}
    \!\!-\!2\text{diag}(\!(\!\mathbf{w}\!\circ\!\mathbf{A}_x\mathbf{h}\!)^{\circ2}\!)\!+\!2\mu_p\mathbf{I}_N.
\end{split}
\end{align} The first term $(\mathbf{w}\circ\mathbf{A}_x\mathbf{h})(\mathbf{w}\circ\mathbf{A}_x\mathbf{h})^{\top}$ is a rank-one matrix with the only non-zero eigenvalue $2||\mathbf{w}\circ\mathbf{A}_x\mathbf{h}||^2$. The second matrix is a diagonal matrix with eigenvalues $\{-2(w_1[\mathbf{A}_x\mathbf{h}]_1)^2,\ldots,\\-2(w_N[\mathbf{A}_x\mathbf{h}]_N)^2\}$. The third matrix is also diagonal but with eigenvalues $2\mu_p$. The Hessian is the sum of a rank one matrix and two diagonal matrices. Its eigenvalues are the sum of the eigenvalues of the these matrices. By the semi-definite convexity condition \cite{boyd2004convex}, each of these eigenvalues now must be greater than or equal to zero. The condition in \eqref{mu_p condition} is sufficient since all $w_i \le w_h$ from the constraint set in \eqref{Interpolation main problem}. \hfill\qed

\bibliographystyle{IEEEtran}
\bibliography{strings,refs}
\end{document}